\def\PLUTO{{\sc pluto}}

\def\gsim{\,\lower4pt\hbox{${\buildrel\displaystyle >\over\sim}$}\,}
\def\lsim{\,\lower4pt\hbox{${\buildrel\displaystyle <\over\sim}$}\,}

\newcommand\rs[1]{_\mathrm{#1}}

\documentclass{aa} 
\usepackage{natbib}
\usepackage{graphicx}
\usepackage{txfonts}
\begin{document}
   \title{Radiative accretion shocks along nonuniform stellar magnetic
          fields in classical T Tauri stars}

   \author{S. Orlando\inst{1}
          \and
          R. Bonito\inst{2,}\inst{1}
          \and
          C. Argiroffi\inst{2,}\inst{1}
          \and
          F. Reale\inst{2,}\inst{1}
          \and
          G. Peres\inst{2,}\inst{1}
          \and
          M. Miceli\inst{1}
          \and
          T. Matsakos\inst{3,4,5}
          \and
          C. Stehl\'e\inst{5}
          \and \\
          L. Ibgui\inst{5}
          \and
          L. de Sa\inst{4,5}
          \and
          J.P. Chi\`eze\inst{4}
          \and
          T. Lanz\inst{6}
          }

   \institute{INAF-Osservatorio Astronomico di Palermo,
           Piazza del Parlamento, 1 , 90134, Palermo, Italy
           \email{orlando@astropa.inaf.it}
         \and
           Dip. di Fisica e Chimica, Universit\`a degli Studi di Palermo,
           Piazza del Parlamento, 1, 90134, Palermo, Italy
         \and
           CEA, IRAMIS, Service Photons, Atomes et Mol\'ecules, 91191
           Gif-sur-Yvette, France
         \and
           Laboratoire AIM, CEA/DSM - CNRS - Universit\'e Paris Diderot,
           IRFU/SAp, 91191 Gif-sur-Yvette, France
         \and
           LERMA, Observatoire de Paris, Universit\'e Pierre et Marie
           Curie, CNRS, 5 place J. Janssen, 92195 Meudon, France
         \and
           Laboratoire Lagrange, Universit\'e de Nice-Sophia Antipolis,
           CNRS, Observatoire de la C\^ote d'Azur, 06304 Nice cedex
           4, France
 }

   \date{}

 
  \abstract
   {According to the magnetospheric accretion model, hot spots form on
    the surface of classical T Tauri stars (CTTSs) where accreting
    disk material impacts onto the stellar surface at supersonic velocity,
    generating a shock.}
   {We investigate the dynamics and stability of post-shock plasma
    streaming along nonuniform stellar magnetic fields at the impact
    region of accretion columns. We study how the magnetic field
    configuration and strength determine the structure, geometry, and
    location of the shock-heated plasma.}
   {We model the impact of an accretion stream onto the chromosphere of
    a CTTS by 2D axisymmetric magnetohydrodynamic simulations. Our model
    takes into account the gravity, the radiative cooling, and the
    magnetic-field-oriented thermal conduction (including the effects
    of heat flux saturation). We explore different configurations and
    strengths of the magnetic field.}
   {The structure, stability, and location of the shocked plasma
    strongly depend on the configuration and strength of the magnetic
    field. In the case of weak magnetic fields (plasma $\beta \gsim 1$
    in the post-shock region), a large component of $\vec{B}$ may develop
    perpendicular to the stream at the base of the accretion column,
    limiting the sinking of the shocked plasma into the chromosphere and
    perturbing the overstable shock oscillations induced by radiative
    cooling. An envelope of dense and cold chromospheric material may
    also develop around the shocked column. For strong magnetic fields
    ($\beta < 1$ in the post-shock region close to the chromosphere),
    the field configuration determines the position of the shock and
    its stand-off height. If the field is strongly tapered close to
    the chromosphere, an oblique shock may form well above the stellar
    surface at the height where the plasma $\beta \approx 1$. We find
    that, in general, a nonuniform magnetic field makes the distribution
    of emission measure vs. temperature of the post-shock plasma lower
    than in the case of uniform magnetic field in particular
    at $T > 10^{6}$~K.}
   {The initial strength and configuration of the magnetic field in the
    region of impact of the stream are expected to influence the
    chromospheric absorption and, therefore, the observability of the
    shock-heated plasma in the X-ray band. In addition, the field strength
    and configuration influence also the energy balance of the shocked
    plasma, its emission measure at $T > 10^{6}$~K being lower
    than expected for a uniform field. The above effects contribute in
    underestimating the mass accretion rates derived in the X-ray band.}


   \titlerunning{Accretion shocks along nonuniform stellar magnetic
   fields in CTTSs}

   \maketitle
%

\section{Introduction}

The environment of Classical T Tauri Stars (CTTSs) is rather complex:
a central protostar interacts actively with the rich and dense
magnetized circumstellar system. According to the magnetospheric
accretion scenario (e.g. \citealt{k91a}), magnetic funnels originate
from the circumstellar disk and guide the plasma toward the central
protostar; the accreting plasma is believed to impact the stellar
surface at, approximately, free-fall velocities (up to a few hundred km
s$^{-1}$), generating a shock with temperature of a few millions degrees
(e.g. \citealt{cg98}). Nowadays, there is a growing consensus in the
literature that the soft component of the X-ray emission detected in
CTTSs, originating from dense ($\gsim 10^{11}$~cm$^{-3}$) plasma, is
produced by these accretion shocks (e.g. \citealt{amp07,gns07}). This
scenario is also corroborated by spatially resolved observations of
bright hot impacts by erupted dense fragments falling back on the Sun
and producing high energy emission in the impact region with many
analogies with the stellar accretion (\citealt{reale}).

Current models provide a plausible global picture of the phenomenon
at work. One-dimensional (1D) numerical studies have shown that the
continuous impact of an accretion column onto the stellar chromosphere
leads to the formation of a slab of dense ($n \gsim 10^{11}$~cm$^{-3}$)
and hot ($T \approx 3-5$ MK) plasma undergoing sandpile oscillations
driven by catastrophic cooling (\citealt{kur08, sao08, soa10}). These
models have been able to reproduce the main features of high spectral
resolution X-ray observations of the CTTS MP Mus (\citealt{sao08})
attributed to post-shock plasma (\citealt{amp07}). These models assume
that the plasma moves and transports energy only along magnetic field
lines. This hypothesis is justified if the plasma $\beta \ll 1$ (where
$\beta$ = gas pressure / magnetic pressure) in the shock-heated material.
The stability and dynamics of accretion shocks in cases where the
low-$\beta$ approximation cannot be applied (and, therefore, the 1D
models cannot be used) have been investigated through two-dimensional
(2D) magnetohydrodynamic (MHD) modeling (\citealt{2010A&A...510A..71O};
in the following Paper I; \citealt{matsakos13}). These 2D simulations
show that the accretion dynamics can be rather complex and that, if 1D
simulations are useful for extrapolating precise physical effects, the
global signatures cannot in average be easily extrapolated from them.
The atmosphere around the impact region of the stream can be strongly
perturbed (depending on the plasma $\beta$), leading to important leaks
at the border of the main stream (Paper I). These results have been
supported by observational indications that the shocked plasma heats and
perturbs the surrounding stellar atmosphere, driving stellar material
into surrounding coronal structures (\citealt{bcd10,dbc12}).

Time-dependent models of radiative accretion shocks therefore provide
a convincing theoretical support to the hypothesis that soft X-ray
emission from CTTSs arises from shocks due to the impact of the accretion
columns onto the stellar surface. However, several points still remain
unclear. The most debated is probably the evidence that, in general, the
mass accretion rates derived from X-rays are consistently lower by one
or more orders of magnitude than the corresponding mass accretion rates
derived from UV/optical/NIR observations (e.g. \citealt{cas11}). This
discrepancy may depend on the fraction of the shocked material that
suffers significant absorption from the thick chromosphere, as the shocked
column partially sinks in the chromosphere, and from the accretion stream
itself (\citealt{soa10}). The stream density is expected to play a crucial
role, as it determines the stand-off height of the hot slab generated
by the impact and the amount of sinking of the slab in the chromosphere
(\citealt{soa10}).

Another debated issue is the evidence that the observed coronal activity
is apparently influenced by accretion. In particular, X-ray luminosity of
CTTSs in the phase of active accretion is observed to be systematically
lower than those of Weak-line T Tauri Stars (WTTSs) with no accretion
signatures (e.g. \citealt{nss95}; see also \citealt{def09} and references
therein). Several different explanations have been proposed in the
literature: either the mass accretion modulates the X-ray emission
through the suppression, disruption, or absorption of the coronal
magnetic activity (e.g. \citealt{fdm03, sab04a, pkf05, jca06, gwj07}),
or, conversely, the X-ray emission modulates the accretion through the
photoevaporation of the circumstellar disk (\citealt{def09}). At variance
with the above arguments, \citet{bcd10} have suggested that the accretion
may even enhance the coronal activity around the region of impact.
These authors have reported on observational evidence that the shocked
plasma resulting from the impact heats the surrounding stellar atmosphere
to soft X-ray emitting temperatures; thus they have proposed a model of
``accretion-fed corona'' in which the accretion provides hot plasma
to populate the surrounding magnetic corona in closed (loops) or open
(stellar wind) magnetic field structures (see also \citealt{dbc12}). A
strong coronal activity on the disk can enhance the mass accretion too:
recently \cite{orp11} have shown, through three-dimensional MHD modeling,
that an intense flare occurring close to the accretion disk can strongly
perturb the disk and trigger mass accretion onto the young star.

The stellar magnetic field is known to play an important role in the mass
accretion process and in the evolution of accretion shocks; for instance,
its strength determines the stability and dynamics of the shocks and the
level of perturbation of the surrounding stellar atmosphere by the hot
post-shock plasma (see Paper I). The configuration of the magnetic field
is expected to be relevant in this process too and it may determine the
geometry and location of the post-shock plasma. The latter point can be
crucial for instance to address open issues related to the absorption
of X-rays arising from the post-shock plasma by the optically thick
chromosphere. Simulations of accretion shocks typically assume a uniform
ambient magnetic field. However the field is likely to be nonuniform in
the region of impact of the accretion stream and one wonders whether it
may influence any of the important open issues of the accretion theory.

In this paper, we extend the analysis of Paper I and investigate the
effects of a nonuniform stellar magnetic field on the evolution of
radiative accretion shocks in CTTSs. In particular, we adopt the 2D
MHD model described in Paper I and explore different configurations
and strengths of the ambient magnetic field. We analyze the role of the
nonuniform magnetic field in the dynamics and confinement of the slab
of shock-heated material and how the surrounding stellar atmosphere can
be perturbed by the impact of the accretion stream onto the stellar
surface. In a forthcoming paper (Bonito et al. in preparation), from
the model results presented here, we synthesize the X-ray emission
arising from the shock-heated plasma, taking into account the effects
of absorption from the dense and cold material surrounding the slab,
and investigate the observability of accretion shock features in the
emerging X-ray spectra.

The paper is organized as follows: in Sect.~\ref{sec2} we describe the
MHD model and the numerical setup; in Sect. \ref{sec3} we describe the
results and, finally, we draw our conclusions in Sect. \ref{sec4}.

\section{MHD modeling}
\label{sec2}

We model the final propagation of an accretion stream through the
atmosphere of a CTTS and its impact onto the chromosphere of the star
(see Paper I for details). The fluid is assumed to be fully ionized with a
ratio of specific heats $\gamma = 5/3$ and treated as an ideal MHD plasma
(the magnetic Reynolds number being $\gg 1$; see Paper I). The stream
is modeled by numerically solving the time-dependent MHD equations of
mass, momentum, and energy conservation. The model takes into account the
(nonuniform) stellar magnetic field, the gravity, the radiative losses
from optically thin plasma, and the thermal conduction (including the
effects of heat flux saturation). The radiative losses are derived with
the PINTofALE spectral code \citep{kd00} including the APED V1.3 atomic
line database \citep{Smith2001ApJ}, and assuming metal abundances of
0.2 of the solar values (as deduced from X-ray observations of CTTSs;
\citealt{tgd07}). Since the ambient magnetic field is organized, the
thermal conduction is anisotropic, being highly reduced in the direction
transverse to the field (e.g. \citealt{spi62}). The heat flux saturation
is also included by following the approach of \cite{obr08} that allows for
a smooth transition between the classical and saturated conduction regime.

The model is implemented using \PLUTO\ (\citealt{mbm07}), a modular,
Godunov-type code for astrophysical plasmas, designed to make efficient
use of massively parallel computers using the message-passing interface
(MPI) for interprocessor communications. The MHD equations are solved
using the Harten-Lax-van Leer Discontinuities (HLLD) approximate
Riemann solver that has been proved to be particularly appropriate
to solve isolated discontinuities formed in the MHD system as, for
instance, accretion shocks (\citealt{mk05}). The evolution of the
magnetic field is carried out using the constrained transport method of
\cite{bs99} that maintains the solenoidal condition at machine accuracy.
At variance with Paper I, here the thermal conduction is treated
separately from advection terms through operator splitting and using the
super-time-stepping technique (\citealt{aag96}) which has been proved
to be very effective to speed up explicit time-stepping schemes for
parabolic problems. This approach is particularly useful for high values
of plasma temperature ($T> 10^6$~K, as in our simulations)
because explicit schemes are subject to a rather restrictive stability
condition (i.e. $\Delta t < (\Delta x)^2/(2\eta)$, where $\eta$ is
the maximum diffusion coefficient), and the thermal conduction timescale
$\tau\rs{cond}$ is generally shorter than the dynamical one $\tau\rs{dyn}$
(e.g. \citealt{hc00, obr08}). Optically thin radiative losses are computed
at the temperature of interest, using a table lookup/interpolation
method, and included in the computation in a fractional step formalism,
preserving the $2^{nd}$ time accuracy, as the advection and source steps
are at least of the $2^{nd}$ order accurate (see \citealt{mbm07}).

We explore different configurations and strengths of the stellar
magnetic field with a set of eight 2D MHD simulations, each covering
about 3000~s of physical time. The MHD equations are solved using
cylindrical coordinates in the plane $(r,z)$, assuming axisymmetry; thus
the coordinate system is oriented in such a way that the stellar surface
lies on the $r$ axis and the stream axis is coincident with the $z$
axis. In the present study, we restrict our analysis to accretion
stream impacts able to produce detectable X-ray emission. \cite{soa10}
have shown that X-ray observations preferentially reveal emission from
the impact of low density ($n\rs{str0} \lsim 10^{12}$~cm$^{-3}$) and high
velocity ($|u_{\rm str0}| \gsim 300$~km s$^{-1}$) accretion streams due
to the large absorption of dense post-shock plasma. X-ray observations
show that the above requirements are fitted, for example, in the well
studied young accreting star MP Mus (\citealt{amp07}) that is an ideal
test case for our analysis. Our simulations, therefore, are tuned on this
case, using star and accretion flow parameters derived from optical
and X-ray observations (\citealt{amp07}). The stellar gravity is
calculated by considering the star mass $M=1.2\, M\rs{\odot}$ and the
star radius $R=1.3\, R\rs{\odot}$. The initial conditions represent an
accretion stream with constant plasma density and velocity, propagating
along the $z$ axis through the stellar corona. The initial unperturbed
stellar atmosphere is magneto-static and consists of a hot (temperature
$T \approx 10^6$~K) and tenuous (plasma density $n_{\rm H} \approx
2\times 10^8$~cm$^{-3}$) corona linked by a steep transition region to an
isothermal chromosphere at temperature $T=10^4$~K and $1.4\times 10^9$~cm
thick. The initial distributions of mass density and temperature of the
stellar atmosphere are derived assuming that the plasma is in pressure and
energy equilibrium; we adapted the wind model of \cite{olr96} to calculate
the initial vertical profiles of mass density and temperature from the
base of the transition region ($T = 10^4$ K) to the corona. Initially
the stream is in pressure equilibrium with the stellar corona and has
a circular cross-section with a radius\footnote{Note that the stream
radius can be strongly reduced before the impact (even by a factor of 5)
because of the tapering of the magnetic field close to the chromosphere
(see Sect.~\ref{sec:nonuniform}).} $r\rs{str} = 10^{10}$~cm. The values
of density and velocity of the stream are derived from the analysis
of X-ray spectra of MP Mus (\citealt{amp07}), namely $n_{\rm str0} =
10^{11}$~cm$^{-3}$ and $u_{\rm str0} = -500$~km~s$^{-1}$ at a height
$z =2.1\times 10^{10}$~cm above the stellar surface\footnote{We have
also considered additional simulations with a different plasma density
(see Tab.~\ref{tab1}).}; its temperature is determined by the pressure
balance across the stream lateral boundary.

\begin{table*}
\centering
\caption{Adopted parameters and initial conditions for the MHD models of accretion shocks}
\label{tab1}
\begin{tabular}{lccccccccc}
\hline
\hline
Model    &  $|\vec{B}\rs{lb}|$ & $\beta\rs{slab}^a$ & $n\rs{str0}$ & $u\rs{str0}$
& $r\rs{str}$ & $\dot{M}$ & magnetic field  & $z\rs{dip}$ & Note\\
abbreviation & [G] & & [cm$^{-3}$] & [km s$^{-1}$] & cm & $[M\rs{\odot} yr^{-1}]$ & configuration   & [cm] \\
\hline
B50-D11-Unif    & 50  & 5    & $10^{11}$  & 500 & $10^{10}$ & $5\times 10^{-11}$ & Uniform           & $-$      & reference case   \\
B500-D11-Unif   & 500 & 0.06 & $10^{11}$  & 500 & $10^{10}$ & $5\times 10^{-11}$ & Uniform           & $-$      & reference case \\
B50-D11-Dip1    & 50  & 10   & $10^{11}$  & 500 & $10^{10}$ & $5\times 10^{-11}$ & Dipole            & $-3\times 10^{10}$  & - \\
B500-D10.7-Dip1 & 500 & 0.05 & $5\times 10^{10}$  & 500 & $10^{10}$ & $2.5\times 10^{-11}$ & Dipole  & $-3\times 10^{10}$  & - \\
B500-D11-Dip1   & 500 & 0.2  & $10^{11}$  & 500 & $10^{10}$ & $5\times 10^{-11}$ & Dipole            & $-3\times 10^{10}$  & - \\
B50-D11-Dip2    & 50  & 100  & $10^{11}$  & 500 & $10^{10}$ & $5\times 10^{-11}$ & Dipole            & $-10^{10}$          & - \\
B500-D10.7-Dip2 & 500 & 10$^b$ & $5\times 10^{10}$  & 500 & $10^{10}$ & $2.5\times 10^{-11}$ & Dipole  & $-10^{10}$        & - \\
B500-D11-Dip2   & 500 & 10$^b$ & $10^{11}$  & 500 & $10^{10}$ & $5\times 10^{-11}$ & Dipole            & $-10^{10}$        & - \\
\hline
\hline
\end{tabular}\\
\medskip
\flushleft{
$^a$ $\beta\rs{slab}$ is an outcome of the simulation and is not set as an initial
parameter.}\\
$^b$ In runs B500-D10.7-Dip2 and B500-D11-Dip2, $\beta\rs{slab}$ is
measured in the slab generated by the oblique shock (see
Sect.~\ref{sec:nonuniform}).
\end{table*}

\begin{figure}[!t]
  \centering
  \includegraphics[width=8.cm]{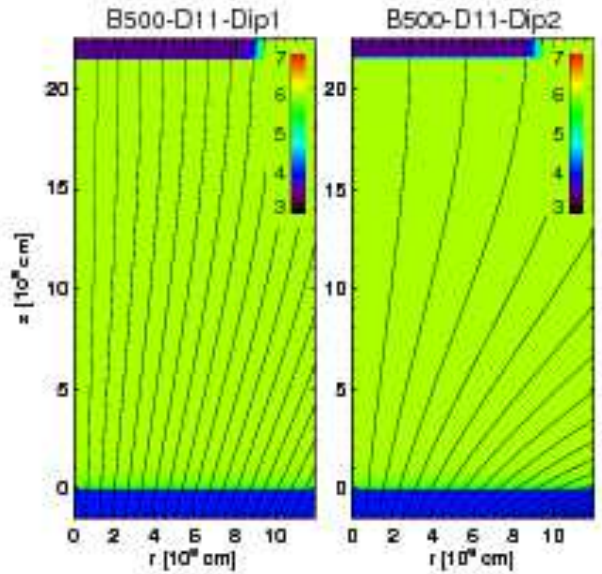}
  \caption{Temperature distribution in the $(r, z)$ plane in log
    scale in the simulations B500-D11-Dip1 and B500-D11-Dip2 at the
    initial conditions ($t=0$). The initial position of the transition
    region between the chromosphere and the corona is at $z = 0$. The
    initial stellar magnetic field is oriented, on the average, along the
    $z$ axis and tapers close to the chromosphere; the black lines mark
    magnetic field lines.}
  \label{fig1}
\end{figure}

The initial stellar magnetic field is assumed to be either uniform and
perpendicular to the stellar surface or characterized by substantial
tapering close to the chromosphere as expected by analogy with
magnetic loop structures (see Fig.~\ref{fig1}). The uniform magnetic
field configuration is considered as a reference, being that used
in Paper I. The nonuniform configuration is realized by assuming
two identical magnetic dipoles both lying on the $z$ axis ($r=0$)
and oriented parallel to it. The first dipole is located either at
$z\rs{dip}=-3\times 10^{10}$~cm or at $z\rs{dip}=-10^{10}$~cm (see models
B500-D11-Dip1 and B500-D11-Dip2 in Fig.~\ref{fig1}); the second dipole
is located specularly with respect to the upper boundary. This idealized
configuration ensures magnetostaticity of the nonuniform field and the
magnetic field oriented parallel to the $z$ axis at the upper boundary
where we impose the inflow of the accretion stream. Such a magnetic field
configuration determines a component of the field perpendicular to the
flow in proximity of the chromosphere. This component may: 1) limit the
overstable shock oscillations, the Lorentz force being not affected by
cooling processes at variance with the force due to gas pressure (e.g.
\citealt{td93}) and 2) limit the sinking of the shocked column in the
chromosphere due to magnetic tension that is expected to sustain the
hot slab. Both effects may influence the observability of the shocked
plasma in the X-ray band.  

From spectrometric (e.g. \citealt{j07, 2011ApJ...729...83Y})
and spectropolarimetric (e.g. BP Tau, AA Tau, V2129 Oph, TW Hya, GQ Lup;
\citealt{2008MNRAS.386.1234D, 2010MNRAS.409.1347D, 2011MNRAS.412.2454D,
2011MNRAS.417..472D, 2012MNRAS.425.2948D}) observations, the inferred
magnetic field strength $|\vec{B}|$ of CTTSs is on the order of few
kG. Given the stream density adopted in our simulations ($n_{\rm str0}
= 10^{11}$~cm$^{-3}$), we considered cases with a minimum value of
magnetic field strength for which $\beta < 1$ in the shocked column
(namely $|\vec{B}| = 500$~G). For larger values of $|\vec{B}|$ the
evolution of the shocked plasma is not expected to change considerably,
$\vec{B}$ behaving rigidly for $\beta < 1$. In all these cases, the
magnetic field is able to confine and channel the shocked plasma.
To investigate shocked flows only partially confined by the magnetic
field, and thus perturbing the surrounding magnetic structures (see,
for instance, Paper I), we considered additional simulations with $\beta$
close to or slightly larger then unity in the shocked column; for $n_{\rm
str0} = 10^{11}$~cm$^{-3}$, this is obtained with $|\vec{B}| = 50$~G
close to the chromosphere. Although this magnetic field is quite low
for a CTTS, we can investigate the dynamics of the stream impact when
the shocked plasma is poorly confined because of a higher densitiy of
the stream and/or because of a weaker field in the impact region.

Additional simulations with a different stream density $n_{\rm str0}$
are also considered. A summary of all the simulations discussed in this
paper is given in Table~\ref{tab1} where: $|\vec{B}\rs{lb}|$ is the
initial magnetic field strength at the lower boundary, $\beta\rs{slab}$
is the average plasma $\beta$ in the hot slab\footnote{This
parameter is an outcome of the simulation and is not set as an initial
parameter.}, $n_{\rm str0}$, $u_{\rm str0}$, and $r\rs{str}$ are the
density, velocity, and radius of the stream at a height $z =2.1\times
10^{10}$~cm, $\dot{M}$ is the mass accretion rate, and $z\rs{dip}$
is the position of the dipole on the $z$ axis.

In all the simulations, the 2D cylindrical $(r,z)$ mesh extends between 0
and $1.2 \times 10^{10}$~cm in the $r$ direction and between $-1.4\times
10^9$~cm and $2.26 \times 10^{10}$~cm in the $z$ direction; the transition
region between the chromosphere and the corona is located at $z=0$~cm. The
radial coordinate $r$ has been discretized uniformly with $N\rs{r} = 512$
points, giving a resolution of $\Delta r \approx 2.3\times 10^7$~cm. The
coordinate $z$ has been discretized on a nonuniform grid with the mesh
size increasing with $z$, giving a higher spatial resolution closer
to the stellar chromosphere. The $z$-grid is made of $N\rs{z} = 896$
points and consists of: i) a uniform grid patch with 512 points and
a maximum resolution of $\Delta z \approx 5.4\times 10^6$~cm covering
the chromosphere and the upper stellar atmosphere up to the height of
$\approx 3\times 10^9$~cm, and ii) a stretched grid patch for $z >
3\times 10^9$~cm with the mesh size increasing with $z$ leading to a
minimum resolution of $\Delta z \approx 5\times 10^7$~cm close to the
upper boundary. This nonuniform mesh allows us to describe appropriately
the steep temperature gradient of the transition region and the evolution
of the hot slab of shock-heated material resulting from the impact
of the accretion stream with the stellar chromosphere.  The boundary
conditions are the same adopted in Paper I: axisymmetric boundary
conditions\footnote{Variables are symmetrized across the boundary, and
both radial and angular $\phi$ components of vector fields ($\vec{u},
\vec{B}$) change their sign.} at $r = 0$ (i.e. along the symmetry axis
of the problem), free outflow\footnote{Set zero gradients across the
boundary.} at $r = 1.2 \times 10^{10}$~cm, fixed boundary conditions at
$z = -1.4\times 10^9$~cm (imposing zero material and heat flux across
the boundary), and a constant inflow in the upper boundary at $z =
2.26 \times 10^{10}$ cm.

\section{Results}
\label{sec3}

\subsection{The reference case: uniform magnetic field}
\label{sec:uniform}

The case of uniform ambient magnetic field is considered here as a
reference. As discussed below, the evolution of the accretion shock and of
the post-shock plasma is analogous to that described in Paper I when the
magnetic field is weak ($\beta \gsim 1$ in the post-shock plasma;
run B50-D11-Unif) and to that described by \cite{soa10} in the limit of
strong magnetic field ($\beta \ll 1$; run B500-D11-Unif).

\begin{figure*}[!th]
  \centering
  \includegraphics[width=17.4cm]{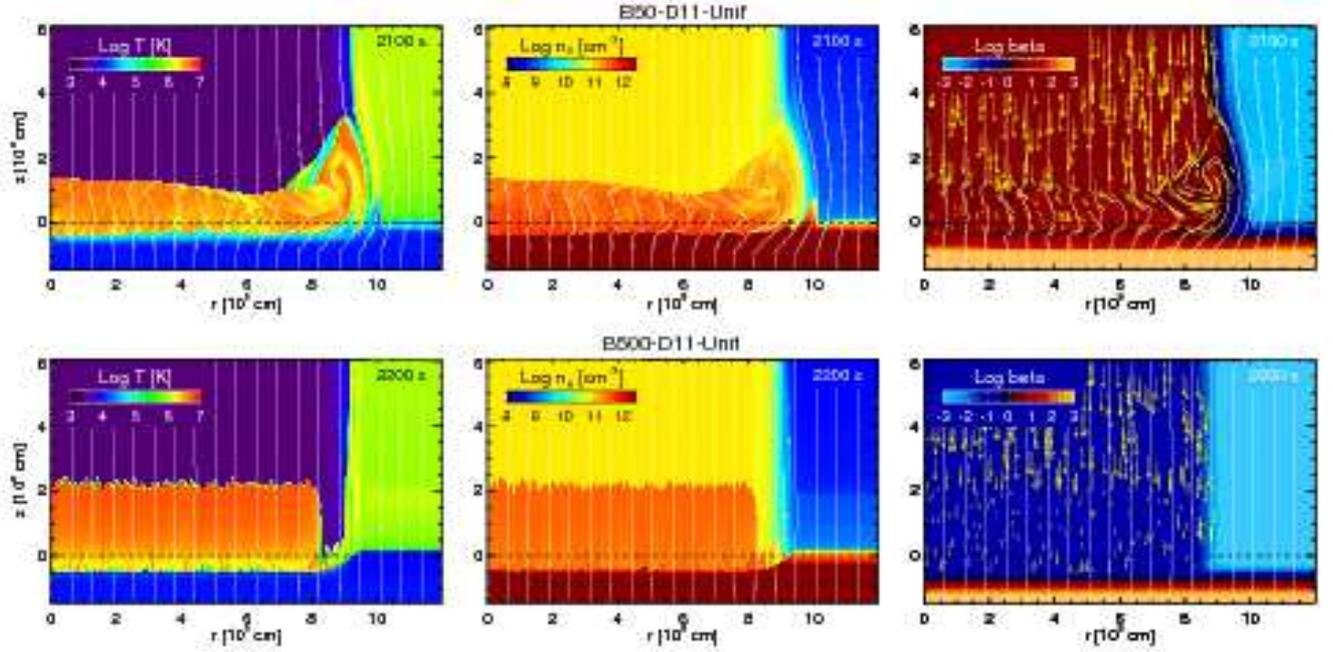}
  \caption{Temperature (left panels), plasma density (center panels), and
     plasma $\beta$ (right panels) distributions in the $(r, z)$ plane
     in log scale in the simulations B50-D11-Unif (upper panels) and
     B500-D11-Unif (lower panels) at the labeled times. The initial
     position of the transition region between the chromosphere and the
     corona is at z = 0 (dashed line). The initial magnetic field is
     uniform and oriented along the $z$ axis; the white lines mark
     magnetic field lines. The yellow arrows in the right panels map
     the velocity field.}
  \label{fig_unif}
\end{figure*}

Figure \ref{fig_unif} shows maps of temperature, density, and $\beta$
for runs B50-D11-Unif and B500-D11-Unif at the labeled times. Movies
showing the complete evolution of 2D spatial distributions of mass density
(on the left) and temperature (on the right) in log scale are provided
as on-line material. In both runs, the accreting plasma flows along
the magnetic field lines and impacts onto the chromosphere at $t\approx
400$~s. After the impact, a hot slab ($T\approx 5$~MK) forms at the base
of the stream; the slab is partially rooted in the chromosphere so that
part of the shocked material is buried under a column of material that
may be optically thick. The slab is thermally unstable and quasi-periodic
oscillations of the shock position are induced by radiative cooling.

In run B50-D11-Unif, $\beta \gsim 1$ in the slab except at
the stream border where $\beta \ll 1$ (see upper right panel in
Fig.~\ref{fig_unif}). The post-shock plasma therefore is confined
efficiently by the magnetic field (no accreting material escapes sideways)
and complex 2D plasma structures form in the slab interior due to thermal
instabilities. The evolution of this simulation is analogous
to that of run By-50 of Paper I and we refer the reader to that case
for more details. In run B500-D11-Unif, the plasma $\beta$ is $\ll 1$
in the slab. As a consequence, the stream results to be structured in
several fibrils, each independent from the others. The strong magnetic
field prevents mass and energy exchanges across magnetic field lines
(see also \citealt{matsakos13}) and the formation of complex 2D plasma
structures in the slab interior as in run B50-D11-Unif. Time-dependent 1D
models as those presented by \cite{soa10} describe each of these fibrils.

\begin{figure}[!ht]
  \centering
  \includegraphics[width=7.5cm]{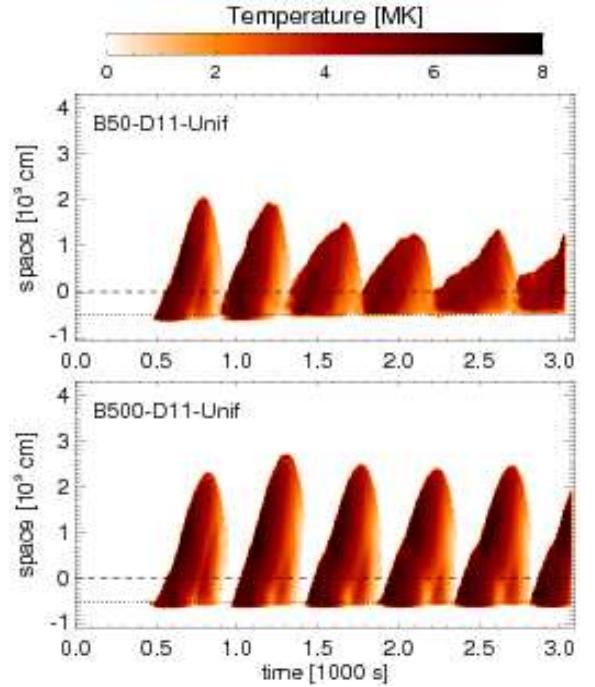}
  \caption{Time-space plots of the emission-measure-weighted temperature
     evolution for runs B50-D11-Unif and B500-D11-Unif. The spatial
     extent of the post-shock plasma in the $z$ direction lies in the
     vertical direction at any time. The dashed line marks the initial
     position of the transition region between the chromosphere and
     the corona; the dotted line marks the minimum sinking of the slab
     into the chromosphere in runs B50-D11-Unif and B500-D11-Unif.}
  \label{fig_diag_unif}
\end{figure}

The global time evolution of runs B50-D11-Unif and B500-D11-Unif can be
analyzed through the time-space plots of the temperature evolution. As
described in Paper I, from the 2D spatial distributions of temperature
and mass density, we first derive the profiles of temperature along the
$z$-axis at each time $t$, by averaging the emission-measure-weighted
temperature along the $r$-axis. Then the time-space plot of temperature
evolution is derived from these profiles. The result is shown in
Fig.~\ref{fig_diag_unif}. In the two cases considered, the hot slab
penetrates the chromosphere down to the position at which the ram
pressure of the post-shock plasma equals the thermal pressure of
the chromosphere. As expected, the slab is not steady and its height
oscillates with a period of $\approx 500$~s due to intense radiative
cooling at the base of the slab (see Paper I for a detailed description
of the system evolution). The maximum height reached is $D\rs{slab}
\approx 3\times 10^{9}$~cm in run B500-D11-Unif and $D\rs{slab}
\approx 2.5\times 10^{9}$~cm in run B50-D11-Unif; in the latter case
the amplitude of the oscillations gradually decreases in the first
1500~s of evolution, and then stabilizes to $D\rs{slab} \approx 2\times
10^{9}$~cm. The oscillations are more regular in run B500-D11-Unif than
in B50-D11-Unif because in the latter case the dynamics of the slab is
perturbed by the evolution of post-shock plasma at the stream border
(see Fig.~\ref{fig_unif}).

\subsection{Nonuniform magnetic field}
\label{sec:nonuniform}

\begin{figure*}[!th]
  \centering
  \includegraphics[width=17.4cm]{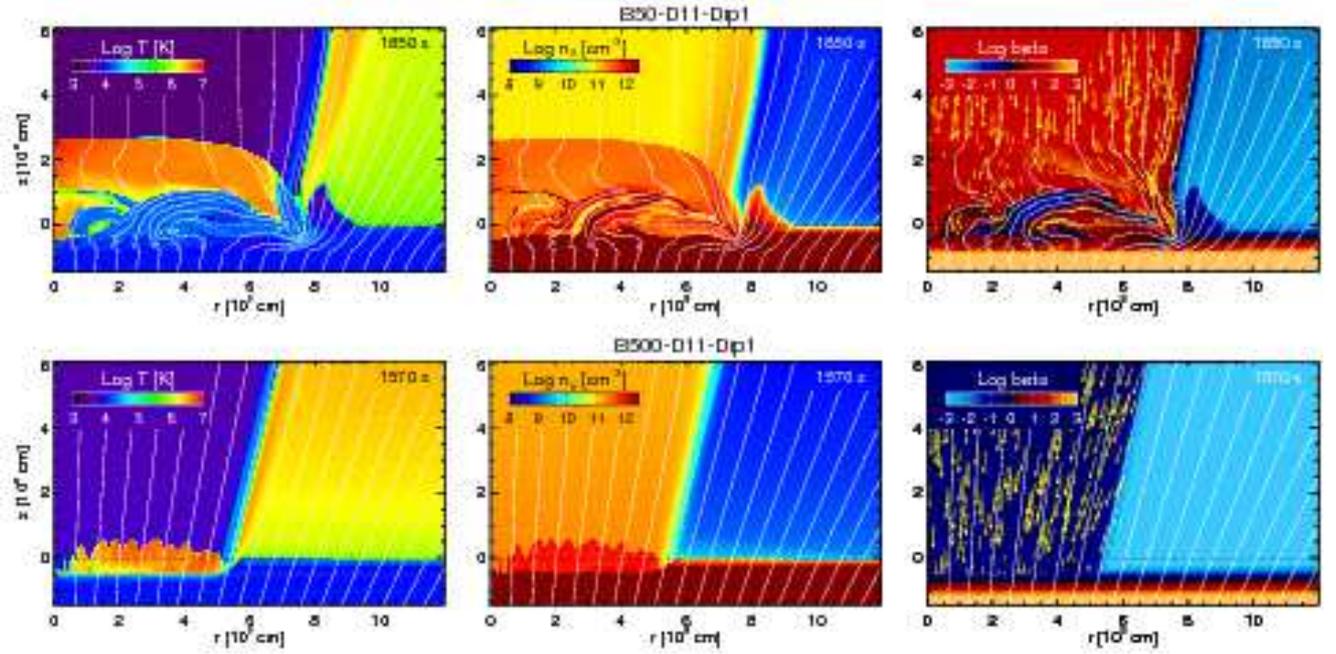}
  \caption{As in Fig.~\ref{fig_unif} for the simulations B50-D11-Dip1
(upper panels) and B500-D11-Dip1 (lower panels).}
  \label{fig_dip1}
\end{figure*}

\begin{figure*}[!th]
  \centering
  \includegraphics[width=17.4cm]{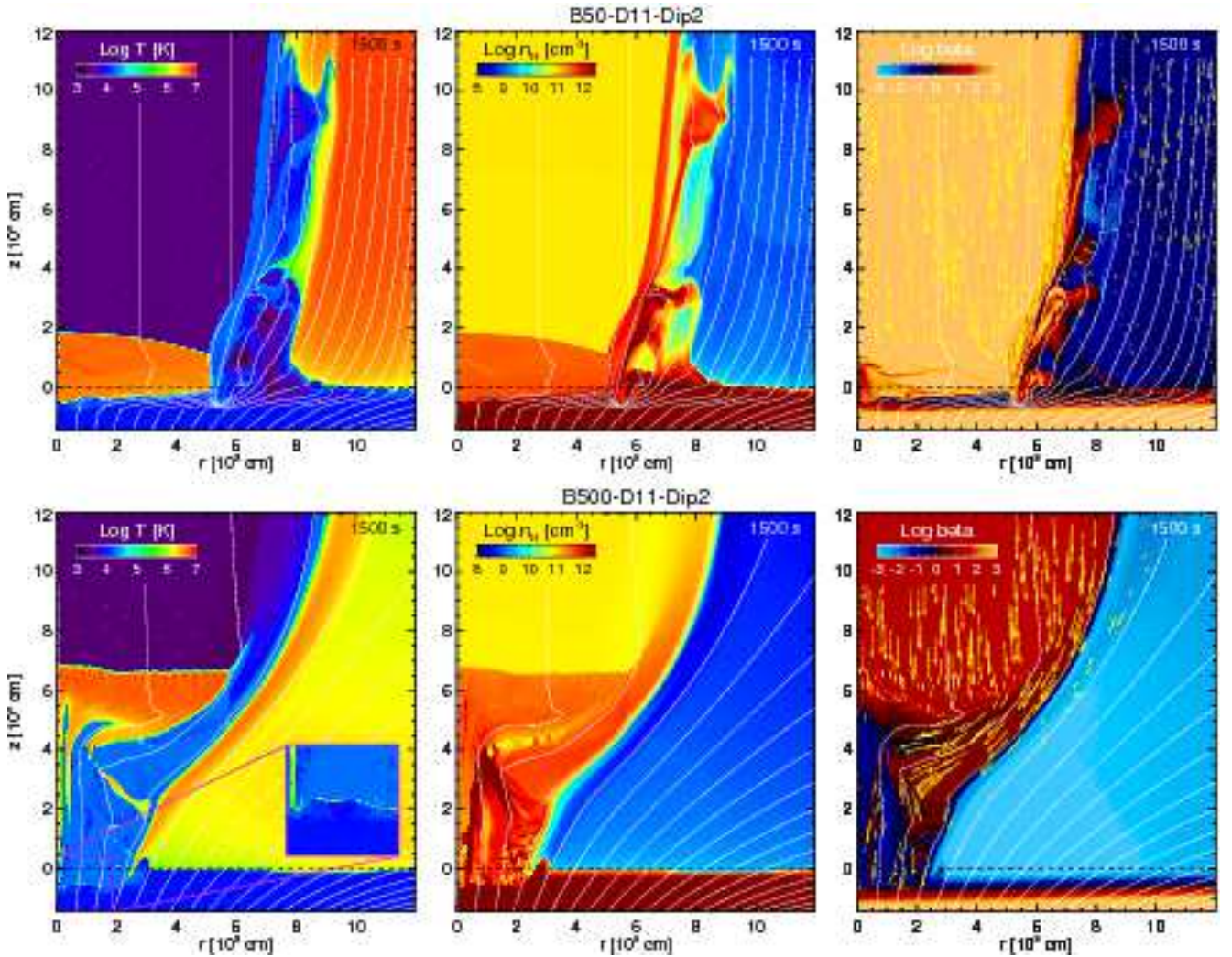}
  \caption{As in Fig.~\ref{fig_unif} for the simulations B50-D11-Dip2
(upper panels) and B500-D11-Dip2 (lower panels). The inset in the lower
left panel shows the shock forming because of the stream impact onto the
chromosphere.}
  \label{fig_dip2}
\end{figure*}

Here we explore how a nonuniform magnetic configuration/topology
influences the dynamics of the shock-heated plasma. In particular, we
investigate the case of $\beta$ increasing from the chromosphere to the
upper atmosphere. Figures~\ref{fig_dip1} and \ref{fig_dip2} show the
spatial distribution of temperature, density, and $\beta$ for different
strengths and configurations of the initial magnetic field. Movies
showing the complete evolution of 2D spatial distributions of mass density
(on the left) and temperature (on the right) in log scale are provided
as on-line material. As in the reference case of uniform magnetic field
(runs B50-D11-Unif and B500-D11-Unif), the stream impact produces a
shock that heats the plasma to few millions degrees (up to $\approx
5$~MK); in all the cases, $\vec{B}$ confines efficiently the post-shock
plasma. At variance with the case of uniform $\vec{B}$, however, the
field tapering makes the stream width decrease progressively (up to a
factor of $\sim 5$ in run B500-D11-Dip2) and consequently the stream
density increase (up to a factor of $\sim 10$ in run B500-D11-Dip2)
while approaching the chromosphere.

When the stellar magnetic field is weak ($\sim 50$~G close to the
chromosphere), the evolution of the shock-heated plasma is in some ways
similar to that described in Sect. \ref{sec:uniform}: a slab of hot
plasma forms at the base of the accretion column and is characterized
by $\beta \gsim 1$ (runs B50-D11-Dip1 and B50-D11-Dip2 in the upper
panels of Figs.~\ref{fig_dip1} and \ref{fig_dip2}). At variance with the
uniform field case, however, a large component of $\vec{B}$ perpendicular
to the stream velocity ($B\rs{r} \approx 10 |\vec{B}\rs{0}|$, where
$\vec{B}\rs{0}$ is the unperturbed magnetic field) develops at the
base of the hot slab (see upper panels of Figs.~\ref{fig_dip1} and
\ref{fig_dip2}). This field component contributes to make the motion of
the shock-heated plasma chaotic and to slightly perturb the overstable
oscillations of the shock. The plasma velocities in the slab range
between 50 and 200~km~s$^{-1}$, namely on the same order of subsonic
turbulent velocities measured from the line profile analysis of Ne~IX
in the CTTS TW Hydrae (\citealt{bcd10}).


\begin{figure}[!t]
  \centering
  \includegraphics[width=8.5cm]{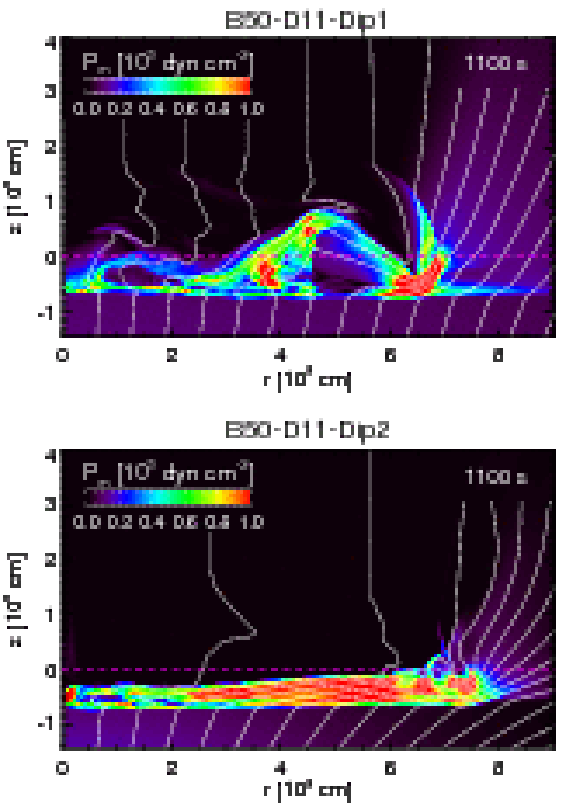}
  \caption{Maps of magnetic pressure $P\rs{m}$ in the $(r, z)$ plane in
     the simulations B50-D11-Dip1 and B50-D11-Dip2 at the labeled
     times. The initial position of the transition region between the
     chromosphere and the corona is at z = 0 (dashed magenta lines). The
     white lines mark magnetic field lines.}
  \label{mag_pres}
\end{figure}

\begin{figure}[!ht]
  \centering
  \includegraphics[width=7.5cm]{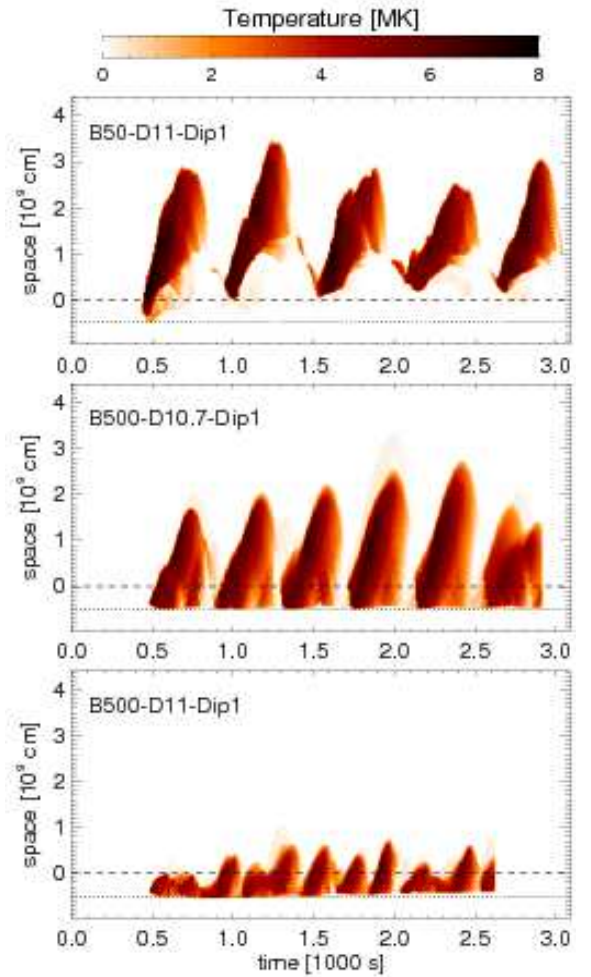}
  \caption{As in Fig.~\ref{fig_diag_unif} for the simulations
  B50-D11-Dip1, B500-D10.7-Dip1, and B500-D11-Dip1.}
  \label{fig_diag_dip1}
\end{figure}

\begin{figure}[!t]
  \centering
  \includegraphics[width=7.5cm]{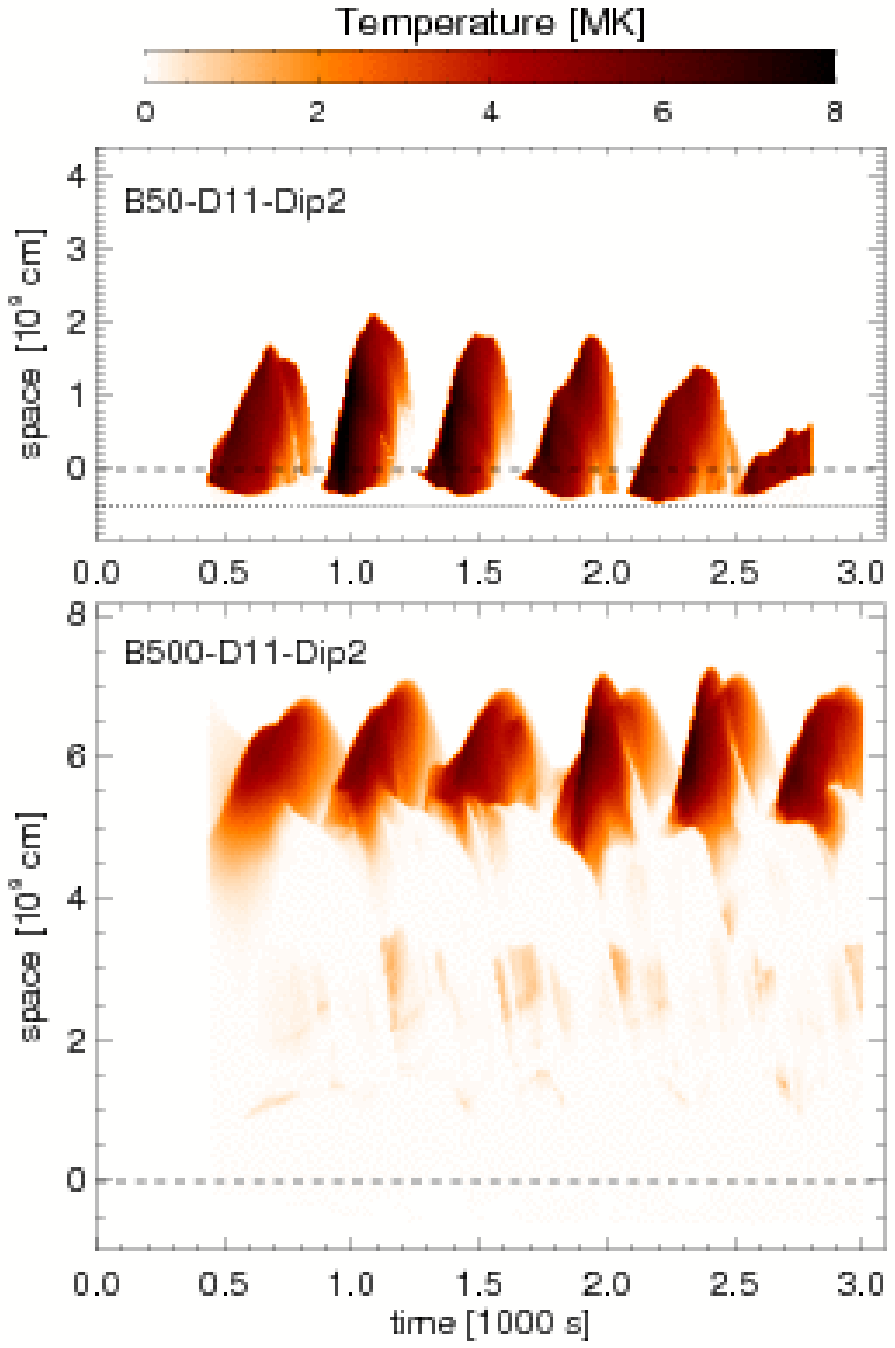}
  \caption{As in Fig.~\ref{fig_diag_unif} for the simulations B50-D11-Dip2
  and B500-D11-Dip2.}
  \label{fig_diag_dip2}
\end{figure}

The perpendicular field component also provides an additional magnetic
pressure at the base of the stream\footnote{The magnetic pressure is
not affected by cooling processes at odds with the gas pressure (see
also \citealt{hp98a}).} (see Fig.~\ref{mag_pres}) which
limits the sinking of the slab into the chromosphere. This is shown in
the upper panels of Figs.~\ref{fig_diag_dip1} and \ref{fig_diag_dip2},
the time-space plots of temperature evolution for runs B50-D11-Dip1
and B50-D11-Dip2, respectively: the slab appears less buried in the
chromosphere (or even above the chromosphere in run B50-D11-Dip1)
than in the uniform field case (see the dotted lines in the figure
representing the minimum sinking of the slab into the chromosphere in
cases with uniform $\vec{B}$). Interestingly \cite{2013ApJS..207....1A}
found no evidence of the post-shock becoming buried in the stellar
chromosphere from the analysis of C~IV line profiles for a sample of
CTTSs. Although our result largely depends on the level of complexity
of the impact region and on the location of the plasma component from
which the emission arises, our model suggests that the bending of
magnetic field lines at the base of the accretion column might also be
considered in the interpretation of the observations.

Since the hot slab is less deep in the chromosphere, the
X-ray emission from the post-shock plasma should be less absorbed by
the optically thick chromosphere in the presence of magnetic field
tapering. However, if the field tapering is large enough (e.g. run
B50-D11-Dip2), the perpendicular component of the magnetic field at the
base of the stream is stronger and more steady and stable. The resulting
excess magnetic pressure, therefore, is able to push chromospheric
material sideways along the magnetic field lines to the upper atmosphere
(see lower panel of Fig.~\ref{mag_pres}); in this case, a sheath of dense
and cold chromospheric material progressively grows around the stream
during the accretion (see upper panels of Fig.~\ref{fig_dip2}). As a
consequence, the shock-heated plasma may be totally enveloped by this
optically thick material providing further absorption of emission from the
impact region. Note that no sheath develops in run B50-D11-Dip1 because
the magnetic field at the base of the stream is much more chaotic, thus
preventing a laminar flow as in run B50-D11-Dip2 (see upper panels of
Figs.~\ref{fig_dip1} and \ref{mag_pres}, and the on-line movie); as
a consequence, the cold and dense plasma at the base of the stream is
not efficiently channeled by the magnetic field outwards to the upper
atmosphere, remaining mostly trapped at the stream base.

When the magnetic field is strong ($\sim 500$~G close to the
chromosphere), $\vec{B}$ is not significantly perturbed by the stream (at
least at the base of the corona) and the plasma flows along the magnetic
field lines before impacting onto the chromosphere (see lower panels
in Figs.~\ref{fig_dip1} and \ref{fig_dip2}). In this case the initial
configuration of $\vec{B}$ turns out to be crucial in determining the
structure, geometry, and location of the shock-heated plasma. If the
magnetic field tapering is small in the region of stream impact (runs
B500-D10.7-Dip1 and B500-D11-Dip1 in Table~\ref{tab1}), the hot slab forms
at the base of the accretion column (as in the uniform field case) and is
structured as a bundle of independent fibrils, each of them describable
in terms of 1D models (see lower panels in Fig.~\ref{fig_dip1}). However,
as already discussed, the field tapering causes the stream to squeeze
and, consequently, to get denser than for a uniform magnetic field
close to the chromosphere. As a result the height of the hot slab,
which depends inversely on the stream density (e.g. \citealt{soa10}),
is much smaller than that in the case of uniform $\vec{B}$ (cfr.
Fig.~\ref{fig_unif}). This is evident by comparing run B500-D11-Dip1
with run B500-D11-Unif (see the time-space plots of temperature
evolution in Figs.~\ref{fig_diag_unif} and \ref{fig_diag_dip1}). We
note that, in run B500-D11-Dip1 only a very small portion of the slab
emerges above the optically thick chromosphere. We expect therefore
that the extent of chromospheric absorption in identical streams can
be very different if the streams impact in regions characterized by
different configurations of the stellar magnetic field. 

Note that, by reducing the initial stream density in order to match
the density value of run B500-D11-Unif close to the chromosphere, the
fibrils have stand-off heights and evolution analogous to those of run
B500-D11-Unif (see center panel in Fig.~\ref{fig_diag_dip1} for run
B500-D10.7-Dip1); however, the mass accretion rate for this stream is
lower than that of run B500-D11-Unif (see Tab.~\ref{tab1}). In
the case of denser accretion streams (leading to higher mass accretion
rates), the evolution of the shocked plasma is expected to be similar to
that of run B500-D11-Dip1 if still $\beta < 1$ in the shocked slab (namely
if $n_{\rm str0} < 5\times 10^{12}$~cm$^{-3}$ in the case of $|\vec{B}|
= 500$~G, assuming a temperature of the hot slab $T \approx 5$~MK):
the plasma flows along the magnetic field lines and forms independent
fibrils. In these cases, however, the shocked slab should be buried
more deeply in the chromosphere, due to the larger ram pressure, and
its thickness should be smaller than in run B500-D11-Dip1 (see Eq. 9
in \citealt{soa10}). On the other hand, if the stream is so dense that
$\beta \gsim 1$ in the shocked slab (this occurs, for instance, if $n_{\rm
str0} > 5\times 10^{12}$~cm$^{-3}$ in the case of $|\vec{B}| = 500$~G),
the evolution of the post-shock plasma is expected to be more similar to
that of run B50-D11-Dip1: the downfalling plasma bends the magnetic field
lines, generating a large component of $\vec{B}$ perpendicular to the
stream at the base of the accretion column. Again, however, the shocked
slab should be deeply buried in the chromosphere due to the high values
of stream density. For these heavy streams, the X-ray emitting shocks
are expected to be hardly visible in X-rays due to strong absorption by
the thick chromosphere and, possibly, by the dense stream itself.


The evolution of the shock-heated material can be significantly
different if $\beta$ varies strongly from the chromosphere to the
upper atmosphere (runs B500-D10.7-Dip2 and B500-D11-Dip2 in
Table~\ref{tab1}). Movies showing the evolution of these runs
are provided as on-line material. An example is shown in the lower
panels of Fig.~\ref{fig_dip2}. In this case, an oblique shock forms
at the height where the plasma $\beta \approx 1$ ($z \approx 6\times
10^9$~cm). There, the stream is well confined by the magnetic field
and the downfalling plasma flows supersonically along the field
lines. Due to the significant field tapering, the slope of the lines
with respect to the $z$ axis gently increases to above a critical
value. At that point an oblique shock forms in order to deflect the
angle of the flow such that the plasma continues to flow parallel
to the field lines. The oblique shock forms close to the stream
border (far from the axis) where the field inclination is larger
(see on-line movies). Then the shock propagates towards the stream
axis where it is reflected. The on-line movies show that the shock
front propagates back and forth between the site where the oblique
shock forms and the symmetry axis, leading to the formation of an
extended hot slab as shown in Fig.~\ref{fig_dip2}. The post-shock
plasma can be locally thermally unstable with overstable shock
oscillations induced by radiative cooling. These oscillations are
evident in the time-space plot of temperature evolution (see lower
panel of Fig.~\ref{fig_diag_dip2}). In the slab, $\beta \gsim 1$
and the evolution of the shock-heated plasma is analogous to that
described in run B50-D11-Unif. It is worth to emphasize that, at
variance with all the other cases investigated in this paper, a
slab of plasma with temperature of a few millions degrees forms
well above the chromosphere at $z \approx 6\times 10^9$~cm (see
lower panels of Figs.~\ref{fig_dip2} and \ref{fig_diag_dip2}). Below
the slab, the plasma flows along the magnetic field lines with
velocities ranging between 200 and 300~km s$^{-1}$, and impacts
onto the chromosphere, producing a second shock (see the inset in
the lower left panel of Fig.~\ref{fig_dip2}). However, given the
high density ($n \approx 10^{12}$~cm$^{-3}$) and low velocity ($u
\approx 300$~km s$^{-1}$) of the plasma before the impact, the
stand-off height of the second shock is very small and the post-shock
plasma results to be fully buried in the chromosphere (thus strongly
absorbed).

Note that, assuming the same strength and configuration of the
magnetic field, the oblique shock forms at different heights in
streams with different densities: the denser the stream, the closer to
the chromosphere the oblique shock forms. On the other hand, \cite{soa10}
have shown that the denser the stream, the smaller the thickness of the
shocked slab, due to the larger efficiency of the radiative losses. In
the case of heavy streams, therefore, we expect that the post-shock
plasma is located very close to the chromosphere (possibly even buried
in it) and with a very small stand-off height. In these cases,
therefore, the absorption by optically thick material is expected to
play an important role.

\begin{figure}[!t]
  \centering
  \includegraphics[width=7.5cm]{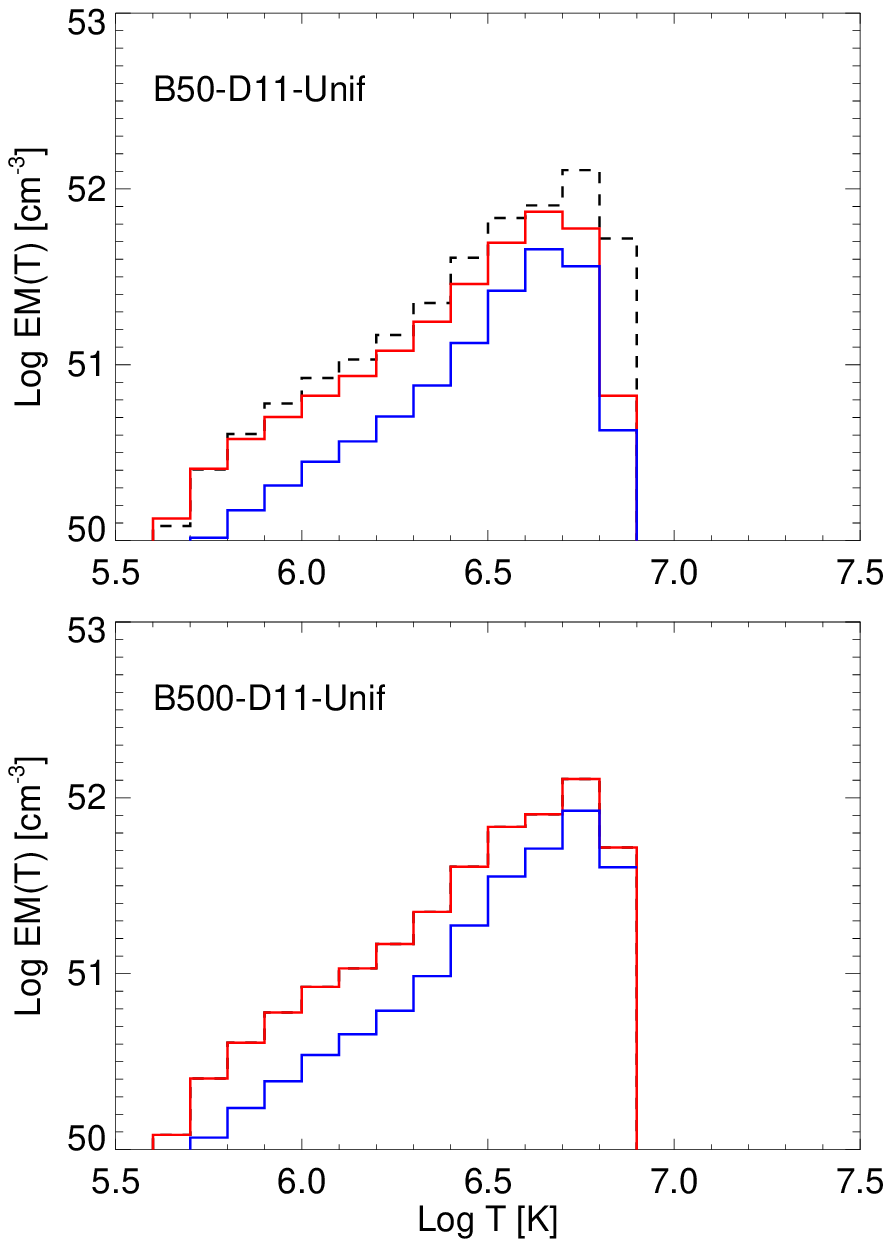}
  \caption{Distributions of emission measure vs. temperature, averaged
    over 3 ks, for runs B50-D11-Unif and B500-D11-Unif. Red (blue) lines
    mark the average EM$(T)$ distributions derived for the whole slab
    (for the portion of the slab emerging above the chromosphere). The
    dashed black line in the upper panel marks the average EM$(T)$
    distribution of the whole slab for run B500-D11-Unif.}
  \label{fig_emt_unif}
\end{figure}

\subsection{Distributions of emission measure vs. temperature}

From the models we derive the distributions of emission measure
vs. temperature EM$(T)$ of the shock-heated material (see Paper I for
details). As shown, for instance, by \cite{amp09}, these distributions
allow straightforwardly to compare the model results with the observations
of accretion shocks in CTTSs. Figure~\ref{fig_emt_unif} shows the
EM$(T)$ distributions averaged over 3~ks (i.e. the total time simulated)
for runs B50-D11-Unif and B500-D11-Unif. The figure shows the EM$(T)$
distributions of the whole slab (red lines) and of the portion of the
slab emerging above the chromosphere (blue lines). The latter
distributions show the fraction of post-shock plasma that is not buried
in the optically thick chromosphere and, therefore, whose X-ray emission
is expected to be less absorbed. In all the cases, we find that the
EM$(T)$ has a peak at $\approx 5$~MK and a shape compatible with those
derived from observations of CTTSs and attributed to plasma heated by
accretion shocks (e.g. \citealt{amp09}).

\begin{figure}[!t]
  \centering
  \includegraphics[width=7.5cm]{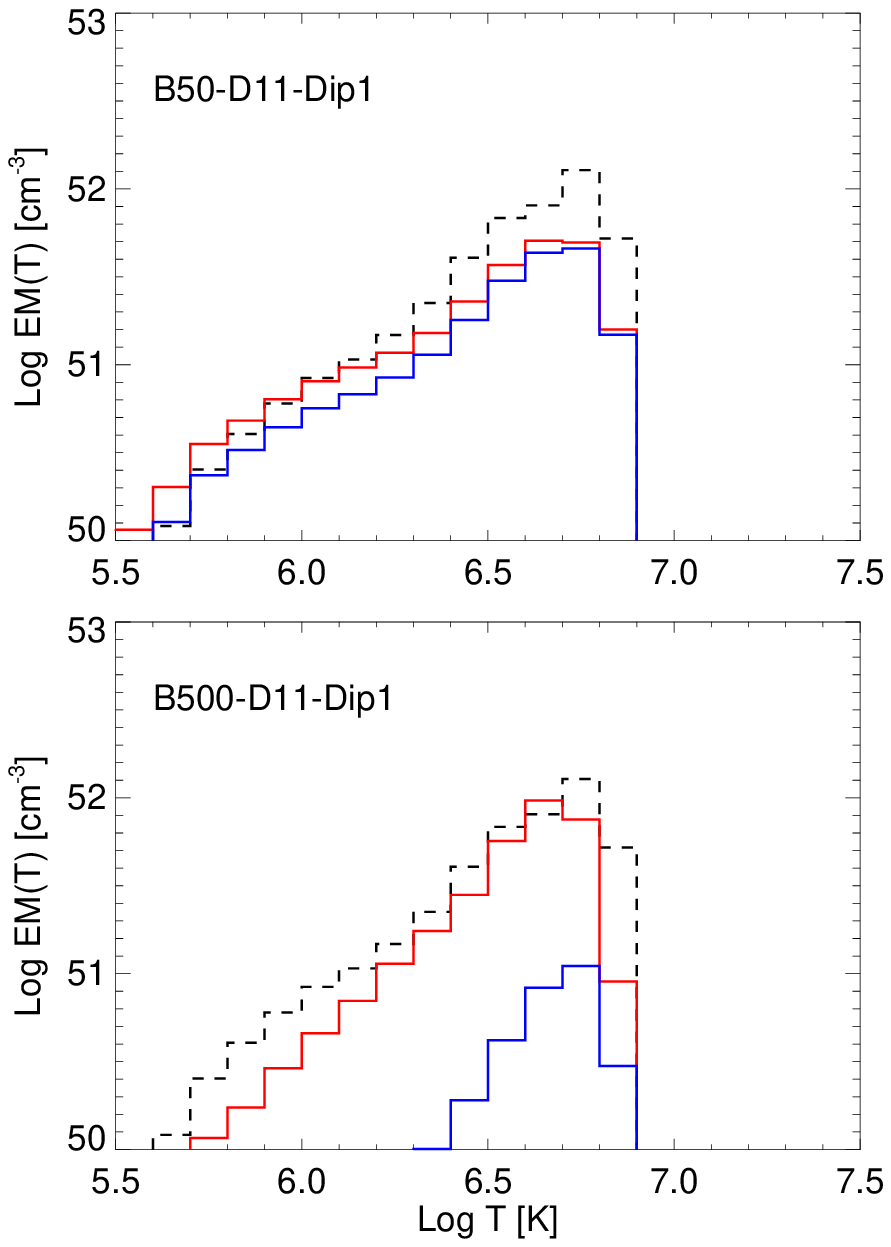}
  \caption{As in Fig.~\ref{fig_emt_unif} for the simulations B50-D11-Dip1
  and B500-D11-Dip1.}
  \label{fig_emt_dip1}
\end{figure}

\begin{figure}[!t]
  \centering
  \includegraphics[width=7.5cm]{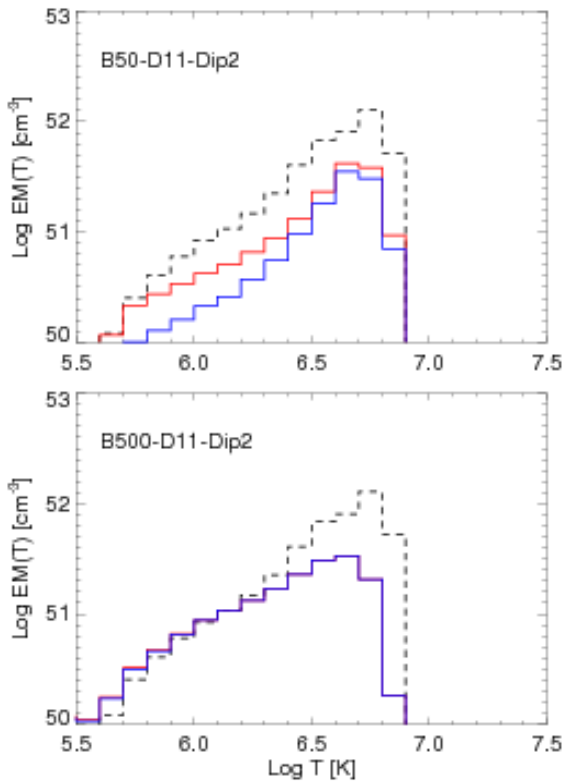}
  \caption{As in Fig.~\ref{fig_emt_unif} for the simulations B50-D11-Dip2
  and B500-D11-Dip2. Note that the blue and red diagrams are almost
  coincident in run B500-D11-Dip2.}
  \label{fig_emt_dip2}
\end{figure}

Figures~\ref{fig_emt_dip1} and \ref{fig_emt_dip2} show the EM$(T)$
distributions averaged over 3~ks of the models with a tapering of
the stellar magnetic field. These distributions have a shape similar
to that obtained for a uniform magnetic field with a peak around
$5-6$~MK. However, in all these cases, the emission measure
EM is lower than that of run B500-D11-Unif (compare red lines with
dashed black lines), although all these models are characterized by
the same mass accretion rate ($\dot{M} \approx 5\times 10^{-11}\,
M\rs{\odot}$~yr$^{-1}$). The difference of EM is the largest at the
highest temperature (above $10^6$~K) where the EM can be significantly
lower than that of run B500-D11-Unif. On the other hand, in general,
the difference decreases at lower temperatures and, in some cases,
the EM can be even slightly higher than that of run B500-D11-Unif at
temperatures below $10^6$~K (see, for instance, runs B50-D11-Dip1 and
B500-D11-Dip2 in Figs.~\ref{fig_emt_dip1} and \ref{fig_emt_dip2}). As
a result, we expect that the emission of the accretion shock is
significanlty reduced in the X-ray band, and only slightly affected in UV
and optical bands. The reduction of emission measure at high temperatures
is larger when the magnetic field is weak (runs B50-D11-Dip1 and
B50-D11-Dip2). In fact, in the presence of a nonuniform magnetic field,
part of the kinetic energy of the stream is spent in bending the magnetic
field lines during the whole evolution (see, for instance, on-line movies
for runs B50-D11-Dip1 and B50-D11-Dip2) and in producing sheaths of dense
and cold chromospheric material enveloping the accretion column (e.g. run
B50-D11-Dip2). In the presence of a strong nonuniform field, the plasma
flow slightly bends the magnetic field lines; however, the field tapering
causes the stream to squeeze and to get denser, resulting in an EM($T$)
distribution peaking at slightly lower temperatures with a steeper
ascending slope (see \citealt{soa10}) as found in run B500-D11-Dip1
(see Fig.~\ref{fig_emt_dip1}). Note that, in run B500-D11-Dip2, the
contribution to the EM$(T)$ distribution comes almost entirely from the
oblique shock at $z \approx 6\times 10^9$~cm rather than from the shock
at the region of stream impact onto the chromosphere (see the inset in
the lower left panel of Fig.~\ref{fig_dip2}).

The amount of post-shock plasma above the chromosphere can be very
different depending on the configuration and strength of the magnetic
field. In the case of weak $\vec{B}$, most of the shock-heated plasma is
above the chromosphere ($\approx 80$\% for run B50-D11-Dip1 and $\approx
75$\% for run B50-D11-Dip2; see upper panels in Figs. \ref{fig_emt_dip1}
and \ref{fig_emt_dip2}), thanks to the large component of $\vec{B}$
perpendicular to the stream velocity developed at the base of the hot
slab which limits the sinking of the shocked column in the chromosphere.
In the case of strong $\vec{B}$, the fraction of post-shock plasma
above the chromosphere strongly depends on the amount of tapering of the
magnetic field, ranging between 10\% for run B500-D11-Dip1 (lower panel
of Fig.~\ref{fig_emt_dip1}) and 99\% for run B500-D11-Dip2 (lower panel
of Fig.~\ref{fig_emt_dip2}). In the former case, the hot slab is buried
in the chromosphere and its stand-off height allows only a small portion
of the post-shock material to emerge. In the latter case, most of the
shock-heated plasma originates from the oblique shock which forms well
above the chromosphere. The configuration and strength of the magnetic
field therefore contribute in determining the geometry and location
of the shocked plasma and are expected to influence the absorption of
the X-ray emitting plasma by the optically thick material surrounding
the hot slab.

It is worth noting that, although the EM$(T)$ distributions
of the portion of the slab emerging above the chromosphere may give
a rough idea of the possible effect of the absorption on the X-ray
emission (assuming that the emission from plasma below the unperturbed
chromosphere is totally absorbed), they do not take into account some
important effects: 1) the absorption by the cold and dense material
from the unperturbed stream above the hot slab and from the perturbed
chromosphere that may surround the slab; 2) the dependence of the
absorption on the wavelength; 3) the point of view from which the impact
region is observed, determining the distribution of thick material along
the line of sight and, therefore, the absorption. To evaluate accurately
the effects of absorption on the emerging X-ray emission, therefore,
it is necessary to synthesize the emission taking into account all the
above points. This issue will be investigated in detail in a forthcoming
paper (Bonito et al. in preparation).

\section{Summary and conclusions}
\label{sec4}

We investigated the stability and dynamics of accretion shocks in CTTSs in
a nonuniform stellar magnetic field, considering different configurations
and strengths of the magnetic field. Our analysis is mainly
focussed on stream impacts able to produce detectable X-ray emission.
We used a 2D axisymmetric MHD model describing the impact of a continuous
accretion stream onto the chromosphere of a CTTS, simultaneously including
the magnetic field, the radiative cooling, and the magnetic-field-oriented
thermal conduction. Our findings lead to the following conclusions.

\begin{itemize}
\item If the plasma $\beta \gsim 1$ where the stream hits the chromosphere
(runs B50-D11-Dip1 and B50-D11-Dip2), the downfalling plasma bends
the magnetic field lines thus generating a large component of $\vec{B}$
perpendicular to the stream at the base of the accretion column. The
perpendicular component limits the sinking of the slab into the
chromosphere and perturbs the overstable oscillations of the shock. As
a result, the fraction of the hot slab emerging above the chromosphere
can be significantly larger than for a uniform magnetic field. If the
tapering of the magnetic field close to the chromosphere is large, a
sheath of dense and cold chromospheric material may also envelope the
accretion column and the hot slab (run B50-D11-Dip2).

\item If the magnetic field is strong enough to confine the downfalling
plasma and guide it towars the chromosphere ($\beta < 1$), the
configuration of $\vec{B}$ determines the position of the shock and its
stand-off height (e.g. runs B500-D11-Dip1 and B500-D11-Dip2). In the
case of a small tapering of $\vec{B}$ close to the chromosphere (runs
B500-D11-Dip1 and B500-D10.7-Dip1), a shock develops at the base of the
accretion column. However, because of the decrease of the stream width
(due to the tapering) and consequent increase of stream density, the
stand-off height of the shock and the fraction of the hot slab emerging
above the chromosphere can be much smaller than that for a uniform
magnetic field.

\item If the tapering of the magnetic field is large (runs B500-D11-Dip2
and B500-D10.7-Dip2), an oblique shock may form well above the
chromosphere at the height where the plasma $\beta \approx 1$. As for
a shock generated by the stream impact, the oblique shock can become
thermally unstable with overstable oscillations induced by radiative
cooling.

\item A nonuniform magnetic field makes, in general, the EM$(T)$
distributions at temperatures above $10^5$~K lower than in the case
of a strong uniform field (run B500-D11-Unif). This effect is larger
at higher temperatures ($T>10^6$~K) and when the magnetic field
is weak (runs B50-D11-Dip1 and B50-D11-Dip2). The main reason is that,
in nonuniform fields, part of the kinetic energy of the downfalling stream
is spent to bend the magnetic field lines during the stream impact and to
develop dense and cold structures of chromospheric material that surround
or even envelop (as in run B50-D11-Dip2) the base of the accretion column.

\end{itemize}

We conclude therefore that the initial strength and configuration of
the magnetic field in the region of impact of the stream with the
chromosphere play an important role in determining the structure,
stability, and location of the post-shock plasma. In particular,
they determine the fraction of the hot slab emerging above the
optically thick chromosphere and the distribution of cold and dense
chromospheric material around or enveloping the shocked column. All
these factors are expected to concur in determining the absorption of
the X-ray emitting plasma and, possibly, in underestimating the mass
accretion rates $\dot{M}$ derived from X-ray observations; this issue
will be investigated in detail in a forthcoming paper (Bonito et al. in
preparation). The strength and configuration of the magnetic field are
expected to influence also the energy balance of the post-shock plasma,
the EM at $T> 10^6$~K being in general significantly lower than
expected assuming a uniform magnetic field. On the other hand, the EM
does not differ too much from that in the presence of a uniform $\vec{B}$
at $T < 10^6$~K, and it can be even larger than in cases with uniform
$\vec{B}$ at temperatures around $10^{5.5}$~K (e.g. runs B50-D11-Dip1
and B500-D11-Dip2). As a consequence, the accretion rates $\dot{M}$ derived from
X-ray observations are again expected to be underestimated if one assumes
a uniform $\vec{B}$. The above results may contribute to explain
the discrepancy between $\dot{M}$ derived from X-rays and the
corresponding values derived from UV/optical/NIR observations in CTTSs
(e.g. \citealt{cas11}).

\begin{acknowledgements}
We thank the referee for constructive and helpful criticism.
\PLUTO\ is developed at the Turin Astronomical Observatory in
collaboration with the Department of General Physics of the Turin
University. We acknowledge the CINECA Award N. HP10BG6HA5,2012
for the availability of high performance computing resources and
support. We acknowledge the computer resources, technical expertise and
assistance provided by the Red Espa\~nola de Supercomputaci\'on (award
N. AECT-2012-2-0001). Additional computations were carried out at the
SCAN\footnote{http://www.astropa.unipa.it/progetti\_ricerca/HPC/index.html}
(Sistema di Calcolo per l'Astrofisica Numerica) facility for high
performance computing at INAF -- Osservatorio Astronomico di Palermo.
TM, CS, LI, LdS, JPC, and TL acknowledge the support of french ANR under
grant 08-BLAN-0263-07.
\end{acknowledgements}

\bibliographystyle{aa}
\bibliography{biblio}

\begin{thebibliography}{43}
\expandafter\ifx\csname natexlab\endcsname\relax\def\natexlab#1{#1}\fi

\bibitem[{{Alexiades} {et~al.}(1996){Alexiades}, {Amiez}, \& {Gremaud}}]{aag96}
{Alexiades}, V., {Amiez}, G., \& {Gremaud}, P.~A. 1996, Communications in
  Numerical Methods in Engineering, 12, 31

\bibitem[{{Ardila} {et~al.}(2013){Ardila}, {Herczeg}, {Gregory}, {Ingleby},
  {France}, {Brown}, {Edwards}, {Johns-Krull}, {Linsky}, {Yang}, {Valenti},
  {Abgrall}, {Alexander}, {Bergin}, {Bethell}, {Brown}, {Calvet}, {Espaillat},
  {Hillenbrand}, {Hussain}, {Roueff}, {Schindhelm}, \&
  {Walter}}]{2013ApJS..207....1A}
{Ardila}, D.~R., {Herczeg}, G.~J., {Gregory}, S.~G., {et~al.} 2013, \apjs, 207,
  1

\bibitem[{{Argiroffi} {et~al.}(2007){Argiroffi}, {Maggio}, \& {Peres}}]{amp07}
{Argiroffi}, C., {Maggio}, A., \& {Peres}, G. 2007, \aap, 465, L5

\bibitem[{{Argiroffi} {et~al.}(2009){Argiroffi}, {Maggio}, {Peres}, {Drake},
  {L{\'o}pez-Santiago}, {Sciortino}, \& {Stelzer}}]{amp09}
{Argiroffi}, C., {Maggio}, A., {Peres}, G., {et~al.} 2009, \aap, 507, 939

\bibitem[{{Balsara} \& {Spicer}(1999)}]{bs99}
{Balsara}, D.~S. \& {Spicer}, D.~S. 1999, Journal of Computational Physics,
  149, 270

\bibitem[{{Brickhouse} {et~al.}(2010){Brickhouse}, {Cranmer}, {Dupree}, {Luna},
  \& {Wolk}}]{bcd10}
{Brickhouse}, N.~S., {Cranmer}, S.~R., {Dupree}, A.~K., {Luna}, G.~J.~M., \&
  {Wolk}, S. 2010, \apj, 710, 1835

\bibitem[{{Calvet} \& {Gullbring}(1998)}]{cg98}
{Calvet}, N. \& {Gullbring}, E. 1998, \apj, 509, 802

\bibitem[{{Curran} {et~al.}(2011){Curran}, {Argiroffi}, {Sacco}, {Orlando},
  {Peres}, {Reale}, \& {Maggio}}]{cas11}
{Curran}, R.~L., {Argiroffi}, C., {Sacco}, G.~G., {et~al.} 2011, \aap, 526,
  A104

\bibitem[{{Donati} {et~al.}(2011{\natexlab{a}}){Donati}, {Bouvier}, {Walter},
  {Gregory}, {Skelly}, {Hussain}, {Flaccomio}, {Argiroffi}, {Grankin},
  {Jardine}, {M{\'e}nard}, {Dougados}, \& {Romanova}}]{2011MNRAS.412.2454D}
{Donati}, J.-F., {Bouvier}, J., {Walter}, F.~M., {et~al.} 2011{\natexlab{a}},
  \mnras, 412, 2454

\bibitem[{{Donati} {et~al.}(2011{\natexlab{b}}){Donati}, {Gregory}, {Alencar},
  {Bouvier}, {Hussain}, {Skelly}, {Dougados}, {Jardine}, {M{\'e}nard},
  {Romanova}, \& {Unruh}}]{2011MNRAS.417..472D}
{Donati}, J.-F., {Gregory}, S.~G., {Alencar}, S.~H.~P., {et~al.}
  2011{\natexlab{b}}, \mnras, 417, 472

\bibitem[{{Donati} {et~al.}(2012){Donati}, {Gregory}, {Alencar}, {Hussain},
  {Bouvier}, {Dougados}, {Jardine}, {M{\'e}nard}, \&
  {Romanova}}]{2012MNRAS.425.2948D}
{Donati}, J.-F., {Gregory}, S.~G., {Alencar}, S.~H.~P., {et~al.} 2012, \mnras,
  425, 2948

\bibitem[{{Donati} {et~al.}(2008){Donati}, {Jardine}, {Gregory}, {Petit},
  {Paletou}, {Bouvier}, {Dougados}, {M{\'e}nard}, {Collier Cameron}, {Harries},
  {Hussain}, {Unruh}, {Morin}, {Marsden}, {Manset}, {Auri{\`e}re}, {Catala}, \&
  {Alecian}}]{2008MNRAS.386.1234D}
{Donati}, J.-F., {Jardine}, M.~M., {Gregory}, S.~G., {et~al.} 2008, \mnras,
  386, 1234

\bibitem[{{Donati} {et~al.}(2010){Donati}, {Skelly}, {Bouvier}, {Gregory},
  {Grankin}, {Jardine}, {Hussain}, {M{\'e}nard}, {Dougados}, {Unruh},
  {Mohanty}, {Auri{\`e}re}, {Morin}, {Far{\`e}s}, \& {MAPP
  Collaboration}}]{2010MNRAS.409.1347D}
{Donati}, J.-F., {Skelly}, M.~B., {Bouvier}, J., {et~al.} 2010, \mnras, 409,
  1347

\bibitem[{{Drake} {et~al.}(2009){Drake}, {Ercolano}, {Flaccomio}, \&
  {Micela}}]{def09}
{Drake}, J.~J., {Ercolano}, B., {Flaccomio}, E., \& {Micela}, G. 2009, \apjl,
  699, L35

\bibitem[{{Dupree} {et~al.}(2012){Dupree}, {Brickhouse}, {Cranmer}, {Luna},
  {Schneider}, {Bessell}, {Bonanos}, {Crause}, {Lawson}, {Mallik}, \&
  {Schuler}}]{dbc12}
{Dupree}, A.~K., {Brickhouse}, N.~S., {Cranmer}, S.~R., {et~al.} 2012, \apj,
  750, 73

\bibitem[{{Flaccomio} {et~al.}(2003){Flaccomio}, {Damiani}, {Micela},
  {Sciortino}, {Harnden}, {Murray}, \& {Wolk}}]{fdm03}
{Flaccomio}, E., {Damiani}, F., {Micela}, G., {et~al.} 2003, \apj, 582, 398

\bibitem[{{Gregory} {et~al.}(2007){Gregory}, {Wood}, \& {Jardine}}]{gwj07}
{Gregory}, S.~G., {Wood}, K., \& {Jardine}, M. 2007, \mnras, 379, L35

\bibitem[{{G{\"u}nther} {et~al.}(2007){G{\"u}nther}, {Schmitt}, {Robrade}, \&
  {Liefke}}]{gns07}
{G{\"u}nther}, H.~M., {Schmitt}, J.~H.~M.~M., {Robrade}, J., \& {Liefke}, C.
  2007, \aap, 466, 1111

\bibitem[{{Hujeirat} \& {Camenzind}(2000)}]{hc00}
{Hujeirat}, A. \& {Camenzind}, M. 2000, \aap, 362, L41

\bibitem[{{Hujeirat} \& {Papaloizou}(1998)}]{hp98a}
{Hujeirat}, A. \& {Papaloizou}, J.~C.~B. 1998, \aap, 340, 593

\bibitem[{{Jardine} {et~al.}(2006){Jardine}, {Collier Cameron}, {Donati},
  {Gregory}, \& {Wood}}]{jca06}
{Jardine}, M., {Collier Cameron}, A., {Donati}, J.-F., {Gregory}, S.~G., \&
  {Wood}, K. 2006, \mnras, 367, 917

\bibitem[{{Johns-Krull}(2007)}]{j07}
{Johns-Krull}, C.~M. 2007, \apj, 664, 975

\bibitem[{{Kashyap} \& {Drake}(2000)}]{kd00}
{Kashyap}, V. \& {Drake}, J.~J. 2000, Bulletin of the Astronomical Society of
  India, 28, 475

\bibitem[{{Koenigl}(1991)}]{k91a}
{Koenigl}, A. 1991, \apjl, 370, L39

\bibitem[{{Koldoba} {et~al.}(2008){Koldoba}, {Ustyugova}, {Romanova}, \&
  {Lovelace}}]{kur08}
{Koldoba}, A.~V., {Ustyugova}, G.~V., {Romanova}, M.~M., \& {Lovelace},
  R.~V.~E. 2008, \mnras, 388, 357

\bibitem[{{Matsakos} {et~al.}(2013){Matsakos}, {Chi{\`e}ze}, {Stehl{\'e}},
  {Gonz{\'a}lez}, {Ibgui}, {de S{\'a}}, {Lanz}, {Orlando}, {Bonito},
  {Argiroffi}, {Reale}, \& {Peres}}]{matsakos13}
{Matsakos}, T., {Chi{\`e}ze}, J.-P., {Stehl{\'e}}, C., {et~al.} 2013, \aap,
  557, A69

\bibitem[{{Mignone} {et~al.}(2007){Mignone}, {Bodo}, {Massaglia}, {Matsakos},
  {Tesileanu}, {Zanni}, \& {Ferrari}}]{mbm07}
{Mignone}, A., {Bodo}, G., {Massaglia}, S., {et~al.} 2007, \apjs, 170, 228

\bibitem[{{Miyoshi} \& {Kusano}(2005)}]{mk05}
{Miyoshi}, T. \& {Kusano}, K. 2005, Journal of Computational Physics, 208, 315

\bibitem[{{Neuhaeuser} {et~al.}(1995){Neuhaeuser}, {Sterzik}, {Schmitt},
  {Wichmann}, \& {Krautter}}]{nss95}
{Neuhaeuser}, R., {Sterzik}, M.~F., {Schmitt}, J.~H.~M.~M., {Wichmann}, R., \&
  {Krautter}, J. 1995, \aap, 297, 391

\bibitem[{{Orlando} {et~al.}(2008){Orlando}, {Bocchino}, {Reale}, {Peres}, \&
  {Pagano}}]{obr08}
{Orlando}, S., {Bocchino}, F., {Reale}, F., {Peres}, G., \& {Pagano}, P. 2008,
  \apj, 678, 274

\bibitem[{{Orlando} {et~al.}(1996){Orlando}, {Lou}, {Rosner}, \&
  {Peres}}]{olr96}
{Orlando}, S., {Lou}, Y.-Q., {Rosner}, R., \& {Peres}, G. 1996, \jgr, 101,
  24443

\bibitem[{{Orlando} {et~al.}(2011){Orlando}, {Reale}, {Peres}, \&
  {Mignone}}]{orp11}
{Orlando}, S., {Reale}, F., {Peres}, G., \& {Mignone}, A. 2011, \mnras, 415,
  3380

\bibitem[{{Orlando} {et~al.}(2010){Orlando}, {Sacco}, {Argiroffi}, {Reale},
  {Peres}, \& {Maggio}}]{2010A&A...510A..71O}
{Orlando}, S., {Sacco}, G.~G., {Argiroffi}, C., {et~al.} 2010, \aap, 510, A71

\bibitem[{{Preibisch} {et~al.}(2005){Preibisch}, {Kim}, {Favata}, {Feigelson},
  {Flaccomio}, {Getman}, {Micela}, {Sciortino}, {Stassun}, {Stelzer}, \&
  {Zinnecker}}]{pkf05}
{Preibisch}, T., {Kim}, Y.-C., {Favata}, F., {et~al.} 2005, \apjs, 160, 401

\bibitem[{{Reale} {et~al.}(2013){Reale}, {Orlando}, {Testa}, {Peres}, {Landi},
  \& {Schrijver}}]{reale}
{Reale}, F., {Orlando}, S., {Testa}, P., {et~al.} 2013, Science, 341, 251

\bibitem[{{Sacco} {et~al.}(2008){Sacco}, {Argiroffi}, {Orlando}, {Maggio},
  {Peres}, \& {Reale}}]{sao08}
{Sacco}, G.~G., {Argiroffi}, C., {Orlando}, S., {et~al.} 2008, \aap, 491, L17

\bibitem[{{Sacco} {et~al.}(2010){Sacco}, {Orlando}, {Argiroffi}, {Maggio},
  {Peres}, {Reale}, \& {Curran}}]{soa10}
{Sacco}, G.~G., {Orlando}, S., {Argiroffi}, C., {et~al.} 2010, \aap, 522, A55

\bibitem[{{Smith} {et~al.}(2001){Smith}, {Brickhouse}, {Liedahl}, \&
  {Raymond}}]{Smith2001ApJ}
{Smith}, R.~K., {Brickhouse}, N.~S., {Liedahl}, D.~A., \& {Raymond}, J.~C.
  2001, \apjl, 556, L91

\bibitem[{{Spitzer}(1962)}]{spi62}
{Spitzer}, L. 1962, {Physics of Fully Ionized Gases} (New York: Interscience,
  1962)

\bibitem[{{Stassun} {et~al.}(2004){Stassun}, {Ardila}, {Barsony}, {Basri}, \&
  {Mathieu}}]{sab04a}
{Stassun}, K.~G., {Ardila}, D.~R., {Barsony}, M., {Basri}, G., \& {Mathieu},
  R.~D. 2004, \aj, 127, 3537

\bibitem[{{Telleschi} {et~al.}(2007){Telleschi}, {G{\"u}del}, {Briggs},
  {Audard}, \& {Scelsi}}]{tgd07}
{Telleschi}, A., {G{\"u}del}, M., {Briggs}, K.~R., {Audard}, M., \& {Scelsi},
  L. 2007, \aap, 468, 443

\bibitem[{{Toth} \& {Draine}(1993)}]{td93}
{Toth}, G. \& {Draine}, B.~T. 1993, \apj, 413, 176

\bibitem[{{Yang} \& {Johns-Krull}(2011)}]{2011ApJ...729...83Y}
{Yang}, H. \& {Johns-Krull}, C.~M. 2011, \apj, 729, 83

\end{thebibliography}

\end{document}